\documentclass[10pt,conference]{IEEEtran}
\usepackage{epsfig,setspace,amsmath,epsf,amssymb,bm,theorem,cite, graphicx, epstopdf, algorithm, algpseudocode,float,color,mathtools}
\usepackage[table,xcdraw]{xcolor}

\newtheorem{theorem}{Theorem}

\newtheorem{remark}{Remark}

\IEEEoverridecommandlockouts
\allowdisplaybreaks

\begin{document}

\title{Rate Distortion Tradeoff in Private Read Update Write in Federated Submodel Learning}

\author{Sajani Vithana \qquad Sennur Ulukus\\
	\normalsize Department of Electrical and Computer Engineering\\
	\normalsize University of Maryland, College Park, MD 20742 \\
	\normalsize \emph{spallego@umd.edu} \qquad \emph{ulukus@umd.edu}}

\maketitle

\begin{abstract}
We investigate the rate distortion tradeoff in private read update write (PRUW) in relation to federated submodel learning (FSL). In FSL a machine learning (ML) model is divided into multiple submodels based on different types of data used for training. Each user only downloads and updates the submodel relevant to its local data. The process of downloading and updating the required submodel while guaranteeing privacy of the submodel index and the values of updates is known as PRUW. In this work, we study how the communication cost of PRUW can be reduced when a pre-determined amount of distortion is allowed in the reading (download) and writing (upload) phases. We characterize the rate distortion tradeoff in PRUW along with a scheme that achieves the lowest communication cost while working under a given distortion budget. 
    
\end{abstract}

\section{Introduction}

The increased use of machine learning (ML) in a wide variety of applications requires a large amount of data owned by various parties in order to train the ML models. This gives rise to processing power limitations in central servers and privacy concerns of users whose data is being used in model training. Federated learning (FL) \cite{FL1,FL2,magazine,Advances} was introduced as a solution to these issues, where each user trains a local model using its local data and shares only the gradients (updates), and not the data itself. This solves user privacy issues to a certain extent, and decentralizes processing power requirements. However, the communication cost of FL is significantly high, since millions of users iteratively communicate the updates and model parameters with the central server. Different methods of reducing the communication cost have been proposed in the literature \cite{sparse1,conv,GGS,fedpaq,billion,recent}. One of them is federated submodel learning (FSL)  where a federated learning (FL) model is divided into multiple submodels based on different types of data used to train the central model. In FSL, a given user downloads (reads) an arbitrary submodel and writes back to the same submodel. This reduces the communication cost and makes the learning process more efficient by eliminating unnecessary downloads/uploads and processing at the users end, since the users only update the set of parameters that can be updated by its local data. However, since the submodels are directly linked to different types of data, the index of the submodel updated by a given user leaks users privacy. Moreover, the updates sent in the writing phase also leak information about users local data according to \cite{InvertingGradients,comprehensive,featureLeakage,DeepLeakage}. Therefore, the updating submodel index and the values of the updates in FSL need to be kept private from the databases in order to guarantee user privacy. This is achieved by private read update write (PRUW) \cite{secureFSL,paper1,rw_jafar,ourICC,pruw,dropout,sparse}, where a user privately downloads (reads) the required submodel and uploads (writes) the corresponding updates while guaranteeing information-theoretic privacy of the submodel index and the values of updates. 

The lowest known reading cost $C_R$ and writing cost $C_W$ achieved by a PRUW scheme is $C_R^*=C_W^*=\frac{2}{1-\frac{2}{N}}$, where $N\geq3$ is the number of non-colluding databases in which the model is stored \cite{ourICC}. This can be reduced further by introducing a given amount of distortion to the downloads and uploads. In FL/FSL, a given amount of distortion, based on the number of users and the nature of the model can be allowed without harming the performance of the model\cite{sparse1,GGS,adaptive,conv}. Note also that the existing methods of reducing the communication cost in FL such as sparsification and quantization already result in distorted uploads and downloads. 

In this work, we introduce a PRUW scheme for FSL that achieves reduced reading and writing costs, compared to $C_R^*$ and $C_W^*$, by introducing a given amount of distortion independently specified for the reading and writing phases. One practical instance of this setting is gradient sparsification in learning with different sparsification rates in the uplink and in the downlink. Typically the sparsification rate in the uplink is lower than that of the downlink due to the limited communication capabilities of users compared to servers. In this case, a pre-determined amount of updates are allowed to be zero in the uplink, and a pre-determined amount of parameters are set to zero in the downlink, resulting in some amount of distortion while achieving a lower communication cost. The proposed scheme guarantees information-theoretic privacy of the updating submodel index and the values of the updates. The scheme also does not reveal the indices of the distorted parameters/updates. The distortion in reading and writing phases is defined based on the Hamming distance between the actual and downloaded/uploaded data. 

The main contributions of this work include, 1) characterization of the rate distortion tradeoff in PRUW, 2) introduction of a PRUW scheme that achieves the lowest known communication cost for a given amount of distortion allowed. 

\section{Problem Formulation}

Consider a PRUW setting with $N$ non-colluding databases storing $M$ independent submodels $\{W_1,\ldots,W_M\}$ of size $L$ each. At each time instance $t$, a user updates an arbitrary submodel without revealing its index or the values of updates. Each submodel consists of symbols from a finite field $\mathbf{F}_q$. Each user downloads the required submodel privately in the reading phase, and uploads the updates privately in the writing phase. Pre-determined amounts of distortion are allowed in the reading and writing phases given by $\tilde{D}_r$ and $\tilde{D}_w$, respectively, in order to reduce the communication cost. 

\emph{Distortion in the reading phase:} A distortion of no more than $\tilde{D}_r$ is allowed in the reading phase, i.e., $D_r\leq \tilde{D}_r$, with
\begin{align}\label{rerror}
    D_{r}=\frac{1}{L}\sum_{i=1}^{L}1_{W_{\theta,i}\neq \hat{W}_{\theta,i}}
\end{align}
where $W_{\theta,i}$, $\hat{W}_{\theta,i}$ are the actual and downloaded versions of the $i$th bit of the required submodel $W_\theta$.

\emph{Distortion in the writing phase:} A distortion of no more than $\tilde{D}_w$ is allowed in the writing phase, i.e., $D_w\leq \tilde{D}_w$, with
\begin{align}\label{werror}
    D_{w}=\frac{1}{L}\sum_{i=1}^{L}1_{\Delta_{\theta,i}\neq \hat{\Delta}_{\theta,i}}
\end{align}
where $\Delta_{\theta,i}$ and $\hat{\Delta}_{\theta,i}$ are the actual and uploaded versions of the $i$th bit of the update to the required submodel.

The goal of this work is to find a scheme that results in the lowest total communication cost under given distortion budgets in the reading and writing phases in the PRUW setting considered. Note that the PRUW setting requires the user required submodel index as well as the values of the updates to be kept private from the databases. 

\emph{Privacy of the submodel index:} No information on the index of the submodel being updated $\theta$ is allowed to leak to any of the databases, i.e., for each $n$,
\begin{align}
    I(\theta^{[t]};Q_n^{[t]},U_n^{[t]}|Q_n^{[1:t-1]},S_n^{[1:t-1]},U_n^{[1:t-1]})=0,
\end{align}
where $Q_n^{[t]}$ and $U_n^{[t]}$ are the query and updates sent by the user to database $n$ at time $t$ in the reading and writing phases and $S_n^{[t]}$ is the storage of database $n$ at time $t$.

\emph{Privacy of the values of updates:} No information on the values of updates is allowed to leak to any of the databases, i.e., for each $\tilde{q}\in\mathbf{F}_q$ and $i\in\{1,\dotsc,L\}$,
\begin{align}
    P(\Delta_{\theta,i}^{[t]}=\tilde{q}|Q_n^{[1:t]},U_n^{[1:t]})=P(\Delta=\tilde{q}),
\end{align}
for each database $n$, where $\Delta_{\theta,i}^{[t]}$ is the update of the $i$th parameter of submodel $\theta$ generated by a given user at time $t$. $P(\Delta=\tilde{q})$, $\tilde{q}\in\mathbf{F}_q$ is the globally known apriori distribution of any given parameter update given by,\footnote{The apriori distribution assumes a uniform distribution on the correctly uploaded updates and zero valued distorted updates.}
\begin{align}
    P(\Delta=\tilde{q})=\begin{cases}
    \tilde{D}_w+\frac{1-\tilde{D}_w}{q}, & \text{if $\tilde{q}=0$},\\
    \frac{1-\tilde{D}_w}{q}, & \text{for each $\tilde{q}\neq 0$}.
    \end{cases}
\end{align}

\emph{Security of submodels:} No information on the submodels is allowed to leak to any of the databases, i.e., for each $n$,
\begin{align}
    I(W_{1:M}^{[t]};S_n^{[t]})=0,
\end{align}
where $W_k^{[t]}$ is the $k$th submodel at time $t$.

In the reading phase, users privately send queries to download the required submodel and in the writing phase, users privately send updates to be added to the existing submodels, i.e., $W_{\theta}^{[t]}=W_{\theta}^{[t-1]}+\Delta_{\theta}^{[t]}$, while ensuring the distortions in the two phases are within the allowed budgets ($\tilde{D}_r$, $\tilde{D}_w$). The reading, writing and total costs are defined as $C_R=\frac{\mathcal{D}}{L}$, $C_W=\frac{\mathcal{U}}{L}$ and $C_T=C_R+C_W$, respectively, where $\mathcal{D}$ is the total number of bits downloaded, $\mathcal{U}$ is the total number of bits uploaded, and $L$ is the size of a submodel.

\section{Main Result}

\begin{theorem}
    For a PRUW setting with $N$ non-colluding databases containing $M$ independent submodels, where $\tilde{D}_r$ and $\tilde{D}_w$ amounts of distortion are allowed in the reading and writing phases, respectively, the following reading and writing costs are achievable,
\begin{align}\label{main}
    (C_R,\ C_W)&=\left(\frac{2}{1-\frac{2}{N}}(1-\tilde{D}_r),\ \frac{2}{1-\frac{2}{N}}(1-\tilde{D}_w)\right).
\end{align}
\end{theorem}

\begin{remark}
The total communication cost decreases linearly with the increasing amounts of distortion allowed in the reading and writing phases.
\end{remark}

\section{Overview of the Proposed Scheme}
The proposed scheme is an extension of the scheme presented in \cite{ourICC} and \cite{dropout}. The scheme in \cite{ourICC} with non-colluding databases considers $\lfloor\frac{N}{2}\rfloor-1$ bits of the required submodel at a time (called subpacketization) and reads from and writes to $\lfloor\frac{N}{2}\rfloor-1$ bits using a single bit in each of the reading and writing phases with no error. In this paper, we consider larger subpackets with more bits, i.e., $\ell\geq\lfloor\frac{N}{2}\rfloor-1$, and correctly read from/write to only $\lfloor\frac{N}{2}\rfloor-1$ selected bits in each subpacket using single bits in the two phases. The rest of the  $\ell-\lfloor\frac{N}{2}\rfloor+1$ bits in each subpacket account for the distortion in each phase, which is maintained under the allowed distortion budgets. The privacy of the updating submodel index as well as the values of updates are preserved in this scheme, while also not revealing the indices of the distorted uploads/downloads. 

The distortion in the proposed scheme is a result of reading and writing zeros (nothing) at a predetermined number of selected parameters in each subpacket based on distortion budget. Thus, the proposed scheme can also be viewed as an efficient private FSL scheme that performs sparsification. \cite{sparse} presents a private FSL scheme with sparsification, where sparsification is performed across subpackets, while this paper performs sparsification within each subpacket.

The proposed scheme consists of the following three tasks: 1) Calculating the optimum reading and writing subpacketizations $\ell_r^*$ and $\ell_w^*$ based on the given distortion budgets $\tilde{D}_r$ and $\tilde{D}_w$. 2) Specifying the scheme, i.e., storage, reading/writing queries and single bit updates, for given values of $\ell_r^*$ and $\ell_w^*$. 3) In cases where the subpacketizations calculated in task 1 are non-integers, the model is divided into two sections and two different integer-valued subpacketizations are assigned to the two sections in such a way that the resulting distortion is within the given budgets. Then, task 2 is independently performed at each of the two sections.

For task 2, note that the scheme in \cite{ourICC} allocates distinct constants $f_i$, $i\in\{1,\dotsc,\ell\}$ to the $i$th bit of each subpacket in all submodels (see \eqref{storage}) in the storage, which makes it possible to combine all parameters/updates in a given subpacket to a single bit in a way that the parameters/updates can be correctly and privately decomposed. However, in this scheme, since there may be two subpacketizations in the two phases, we need to ensure that each subpacket in both phases consists of bits with distinct associated $f_i$s. In order to do this, we associate distinct $f_i$s with each consecutive $\max\{\ell_r^*,\ell_w^*\}$ bits in a cyclic manner so that each subpacket in both phases have distinct $f_i$s. The scheme is explained in detail next.

\section{Proposed Scheme}\label{schm}

The scheme is defined on a single subpacket in each of the two phases, and is applied repeatedly on all subpackets. Since the number of bits correctly downloaded/updated remains constant at $\lfloor\frac{N}{2}\rfloor-1$ for a given $N$, the distortion in a subpacket of size $\ell$ is $\frac{\ell-\lfloor\frac{N}{2}\rfloor+1}{\ell}$. Note that this agrees with the definitions in \eqref{rerror} and \eqref{werror} since the same distortion is resulted by all subpackets. Therefore, the optimum subpacketizations in the two phases, $\ell_r^*$ and $\ell_w^*$, are functions of $\tilde{D}_r$, $\tilde{D}_w$ and $N$, and will be calculated in Section~\ref{opt}. First, we describe the general scheme for any given $\ell_r^*$ and $\ell_w^*$. 

\emph{Storage:} The storage of $y=\max\{\ell_r^*, \ell_w^*\}$ bits of all submodels in database $n$, $n\in\{1,\dotsc,N\}$ is given by,
\begin{align}S_n=
    \begin{bmatrix}
    \frac{1}{f_{1}-\alpha_n}\begin{bmatrix}
    W_{1,1}\\\vdots\\W_{M,1}
    \end{bmatrix}+\sum_{j=0}^{\lfloor\frac{N}{2}\rfloor-1}\alpha_n^j I_{1,j}\\
    \vdots\\
    \frac{1}{f_{y}-\alpha_n}\begin{bmatrix}
    W_{1,y}\\\vdots\\W_{M,y}
    \end{bmatrix}+\sum_{j=0}^{\lfloor\frac{N}{2}\rfloor-1}\alpha_n^j I_{y,j}
    \end{bmatrix},\label{storage}
\end{align}
where $W_{i,j}$ is the $j$th bit of submodel $i$ and the $I$s are random noise vectors of size $M\times1$. The scheme is studied under two cases, 1) $y=\ell_w^*\geq\ell_r^*$, and 2) $y=\ell_r^*>\ell_w^*$.

\subsection{Case 1: $y=\ell_w^*\geq \ell_r^*$}

\begin{figure*}[t]
    \centering
    \includegraphics[scale=0.85]{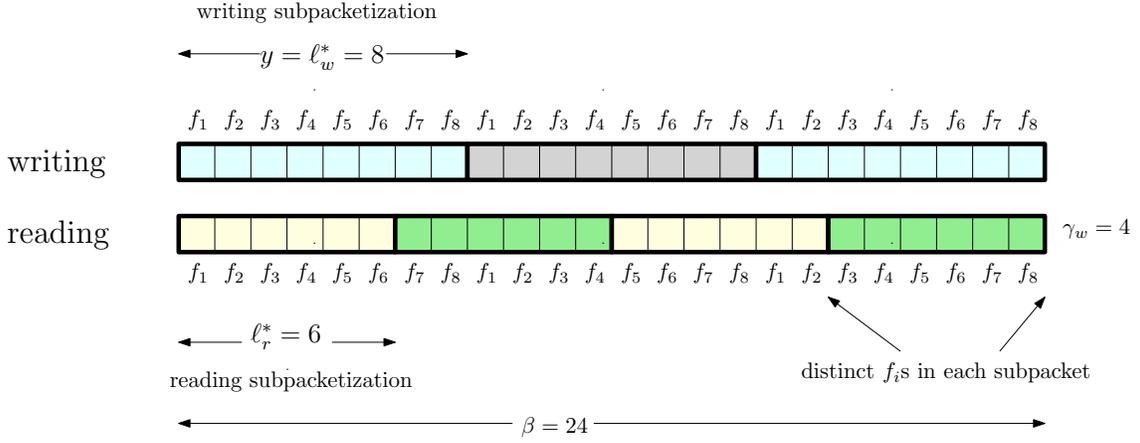}
    \caption{An example setting for case 1.}
    \label{fig1}
\end{figure*}

\emph{Reading phase:} In this case, the user considers subpackets of size $\ell_r^*$ and only downloads $\lfloor\frac{N}{2}\rfloor-1$ bits of each subpacket. Note that each consecutive $y=\ell_w^*$ bits in storage are associated with distinct $f_i$s, which makes each consecutive set of $\ell_r^*$ (reading subpacket size) $f_i$s distinct as well. However, not all reading subpackets have the same $f_i$ allocated to their $i$th bit due to the definition of the storage structure (cyclic allocation of $\ell_w^*$ distinct values of $f_i$). Therefore, we cannot define the reading query on a single subpacket and use it repeatedly, since the reading queries depend on $f_i$s. Thus, we define $\gamma_r=\frac{\text{lcm}\{\ell_r^*,\ell_w^*\}}{\ell_r^*}$ queries to read any $\gamma_r$ consecutive subpackets. Note that the \emph{super subpacket} which consists of any $\gamma_r$ consecutive reading subpackets have the same set of $f_i$s that occur in a cyclic manner in the storage. Therefore, the $\gamma_r$ queries can be defined once on a \emph{super subpacket}, and can be used repeatedly throughout the process. An example setting is given in Fig.~\ref{fig1}, where the reading and writing subpacketizations are given by $\ell_r^*=6$, $\ell_w^*=8$ and the storage structure repeats at every $y=8$ bits. Each square in Figure~\ref{fig1} corresponds to a single bit of all submodels associated with the corresponding value of $f_i$. It shows three consecutive storage/writing subpackets on the top row. The same set of bits are viewed as $\gamma_r=\frac{\text{lcm}\{6,8\}}{6}=4$ reading subpackets, each of size $\ell_r^*=6$ in the bottom row. Note that each reading subpacket contains distinct $f_i$s, which are not the same across the four subpackets. However, it is clear that the structure of the \emph{super subpacket} which contains the four regular subpackets keeps repeating with the same set of $f_i$s in order. The reading phase has the following steps.

The user sends the following queries to database $n$, $n\in\{1,\dotsc,N\}$ to obtain each of the arbitrary sets of $\lfloor\frac{N}{2}\rfloor-1$ bits of each subpacket in each set of $\gamma_r=\frac{\text{lcm}\{\ell_r^*,\ell_w^*\}}{\ell_r^*}$ consecutive, non-overlapping subpackets. Let $J_r^{[s]}$ be the set of $\lfloor\frac{N}{2}\rfloor-1$ parameter indices that are read correctly from subpacket $s$ for $s\in\{1,\dotsc,\gamma_r\}$. The query to download subpacket $s$ is,
\begin{align}\label{query_read2}
        Q_n(s)=\! \begin{bmatrix}
        e_M(\theta)1_{\{1\in J_r^{[s]}\}}\!+\!(f_{g((s-1)\ell_r^*+1)}-\alpha_n)\tilde{Z}_{s,1}\\
        \vdots\\
        e_M(\theta)1_{\{\ell_r^*\in J_r^{[s]}\}}\!+\!(f_{g(s\ell_r^*)}-\alpha_n)\tilde{Z}_{s,\ell_r^*}
    \end{bmatrix}, 
\end{align}
and the corresponding subpacket $s$ is,
\begin{align}
\!\!S_n^{[s]}=\!
   \begin{bmatrix}
    \frac{1}{f_{g((s-1)\ell_r^*+1)}-\alpha_n}\begin{bmatrix}
    W_{1,1}^{[s]}\\\vdots\\W_{M,\ell_r^*}^{[s]}
    \end{bmatrix}\!+\!\sum_{j=0}^{\lfloor\frac{N}{2}\rfloor-1}\alpha_n^j I_{1,j}^{[s]}\\
    \vdots\\
    \frac{1}{f_{g(s\ell_r^*)}-\alpha_n}\begin{bmatrix}
    W_{1,\ell_r^*}^{[s]}\\\vdots\\W_{M,\ell_r^*}^{[s]}
    \end{bmatrix}+\sum_{j=0}^{\lfloor\frac{N}{2}\rfloor-1}\alpha_n^j I_{\ell_r^*,j}^{[s]}
    \end{bmatrix},\label{extend}
\end{align}
where $e_M(\theta)$ is the all zeros vector of size $M\times1$ with a $1$ at the $\theta$th position, $\tilde{Z}$s are random noise vectors of size $M\times1$ and the function $g(\cdot)$ is defined as,
\begin{align}\label{gfunc}
    g(x)=\begin{cases}
    x\!\!\!\!\mod y, & \text{if $x\!\!\!\!\mod y\neq 0$}\\
    y, & \text{if $x\!\!\!\!\mod y=0$}
    \end{cases}
\end{align}

Note that $S_n=[S_n^{[1]},\dotsc, S_n^{[\gamma_r]}]^T$ is the concatenation of $\frac{\text{lcm}\{\ell_r^*,\ell_w^*\}}{y}$ blocks of the form \eqref{storage}. The $\gamma_r$ answers received by database $n$, $n\in\{1,\dotsc,N\}$, are given by,
\begin{align}
    A_n(s)=&(S_n^{[s]})^TQ_n(s),\quad s\in\{1,\dotsc,\gamma_r\}\\
    =&\sum_{i=1}^{\ell_r^*} \left(\frac{1}{f_{g((s-1)\ell_r^*+i)}-\alpha_n}W_{\theta,i}^{[s]}\right)1_{\{i\in J_r^{[s]}\}}\nonumber\\
    &\quad+P_{\alpha_n}(\lfloor\frac{N}{2}\rfloor).
\end{align}
Since $|J_r^{[s]}|=\lfloor\frac{N}{2}\rfloor-1$ for each $s\in\{1,\dotsc,\gamma_r\}$, the required $\lfloor\frac{N}{2}\rfloor-1$ bits of each of the $\gamma_r$ subpackets can be correctly downloaded from the $N$ answers above.

\emph{Writing phase:} Since the subpacketization in the writing phase is $y$, which is the same as the period of the cyclic structure of the storage in \eqref{storage}, a single writing query, specifying the submodel index and the correctly updated bit indices, defined on a single subpacket suffices to repeatedly update all subpackets, as the $f_i$s in all subpackets are identical. The writing query sent to database $n$, $n\in\{1,\dotsc,N\}$, is,
\begin{align}
    \tilde{Q}_n=
    \begin{bmatrix}
        \frac{1}{f_{1}-\alpha_n}e_M(\theta)1_{\{1\in J_w\}}+\hat{Z}_{1}\\
        \vdots\\
        \frac{1}{f_{y}-\alpha_n}e_M(\theta)1_{\{y\in J_w\}}+\hat{Z}_{y}
    \end{bmatrix},
\end{align}
where $J_w$ is the set of indices of the $\lfloor\frac{N}{2}\rfloor-1$ parameters of each subpacket, that are updated correctly and $\hat{Z}$s are random noise vectors of size $M\times1$. Since $\tilde{Q}_n$ is sent only once, the same set of $J_w$ indices will be correctly updated in all subpackets. The user then sends a single bit combined update for each subpacket of the form \eqref{storage} given by,
\begin{align}
    U_n=\sum_{i\in J_w} \tilde{\Delta}_{\theta,i}\prod_{j\in J_w, j\neq i}(f_j-\alpha_n)+\prod_{j\in J_w}(f_j-\alpha_n)z, 
\end{align}
where $\tilde{\Delta}_{\theta,i}=\frac{\Delta_{\theta,i}}{\prod_{j\in J_w, j\neq i}(f_j-f_i)}$ and $z$ is a random noise bit. Each database then calculates the incremental update as,
\begin{align}
    \tilde{U}_n&=U_n\times \tilde{Q}_n\\
    &=\begin{bmatrix}
        \frac{\Delta_{\theta,1}}{f_{1}-\alpha_n}e_M(\theta)1_{\{1\in J_w\}}+P_{\alpha_n}(\lfloor\frac{N}{2}\rfloor-1)\\
        \vdots\\
        \frac{\Delta_{\theta,y}}{f_{y}-\alpha_n}e_M(\theta)1_{\{y\in J_w\}}+P_{\alpha_n}(\lfloor\frac{N}{2}\rfloor-1)
    \end{bmatrix},
\end{align}
which can be directly added to the existing storage in \eqref{storage} to obtain the updated model. The reading and writing costs of case 1 are given by,
\begin{align}
    C_R^{[1]}&=\frac{\gamma_r\times N}{\gamma_r\times \ell_r^*}=\frac{N}{\ell_r^*}, \qquad 
    C_W^{[1]}=\frac{N}{\ell_w^*}\label{c2}.
\end{align}

\subsection{Case 2: $y=\ell_r^*\geq \ell_w^*$}

This case is unlikely to occur in practice in relation to sparsification, since a higher subpacketization implies higher allowed distortion, which essentially means a lower sparsification rate in the downlink compared to the uplink. Typically, it is the server that has a higher communication capacity which makes the downlink sparsification rate larger than that of the uplink, which is contradicting to this case. Due to smaller liklelihood of occurring in practice, space limitations here, and similarities to case 1, we skip the details of the scheme corresponding to case 2. In summary, the storage is the same as \eqref{storage} with $y=\ell_r^*$, and the reading phase is similar to \cite{ourICC} with identity functions specifying the non-zero parameter indices.

The writing phase considers \emph{super subpackets} similar to the reading phase in case 1 containing $\gamma_w=\frac{\text{lcm}\{\ell_r^*,\ell_w^*\}}{\ell_w^*}$ subpackets. The writing queries for each of the $s$, $s\in\{1,\dotsc,\gamma_w\}$ subpackets, which are sent only once, is,
\begin{align}\label{query_write}
        \tilde{Q}_n(s)=\begin{bmatrix}
        \frac{1}{f_{g((s-1)\ell_w^*+1)}-\alpha_n}e_M(\theta)1_{\{1\in J_w^{[s]}\}}+\hat{Z}_{s,1}\\
        \vdots\\
        \frac{1}{f_{g(s\ell_w^*)}-\alpha_n}e_M(\theta)1_{\{\ell_w^*\in J_w^{[s]}\}}+\hat{Z}_{s,\ell_w^*}
    \end{bmatrix},  
\end{align}
with $J_w^{[s]}$ being the indices of the correctly updated parameters of subpacket $s$. The combined update of each subpacket is,
\begin{align}\label{update}
    U_n(s)=&\sum_{i\in J_w^{[s]}} \tilde{\Delta}_{\theta,i}^{[s]}\prod_{j\in J_w^{[s]}, j\neq i}(f_{g((s-1)\ell_w^*+j)}-\alpha_n)\nonumber\\
    &\quad+\prod_{j\in J_w^{[s]}}(f_{g((s-1)\ell_w^*+j)}-\alpha_n)z, 
\end{align}
and the incremental update is calculated by $\tilde{U}_n(s)=U_n(s)\times \tilde{Q}_n(s)$, for each $s\in\{1,\dotsc,\gamma_w\}$, which is directly added to the corresponding $\gamma_w$ subpackets in storage. The reading and writing costs for case 2 are given by,
\begin{align}
    C_R^{[2]}&=\frac{N}{\ell_r^*}, \qquad
    C_W^{[2]}=\frac{\gamma_w \times N}{\gamma_w \times \ell_w^*}=\frac{N}{\ell_w^*}\label{c1}.
\end{align}

\begin{remark}
Note that the cost of sending $Q_n$ and $\tilde{Q}_n$ is not considered in the above writing cost since they are sent only once to each database in the entire PRUW process (i.e., not per subpacket) and the combined cost of $Q_n$ and $\tilde{Q}_n$ given by $\frac{M(\ell_r^*+\text{lcm}\{\ell_r^*,\ell_w^*\})}{L}$ is negligible since $L$ is very large. 
\end{remark}

\subsection{Calculation of Optimum $\ell_r^*$ and $\ell_w^*$ for Given ($\tilde{D}_r$, $\tilde{D}_w$)} \label{opt}

In order to minimize the total communication cost, the user correctly reads from and writes to only $\lfloor\frac{N}{2}\rfloor-1$ out of each of the $\ell_r^*$ and $\ell_w^*$ bits in reading and writing phases, respectively. This results in an error that needs to be kept within the given distortion budgets of $\tilde{D}_r$ and $\tilde{D}_w$. Note from \eqref{c2} and \eqref{c1} that the reading and writing costs follow a symmetric pattern. Therefore, the minimization of $C_T=C_R+C_W$ can be considered as two identical and independent minimizations of $C_R$ and $C_W$, since $\ell_r^*$ and $\ell_w^*$ only depend on $\tilde{D}_r$ and $\tilde{D}_w$, which are independent. Therefore, due to symmetry, we drop the subscripts of $\ell$ and $D$ in the following steps, i.e., we use a generic $\ell$ in place of $\ell_r^*$ and $\ell_w^*$, and similarly a generic $\tilde{D}$ in place of $\tilde{D}_r$ and $\tilde{D}_w$. 

For a subpacketization $\ell=\lfloor\frac{N}{2}\rfloor-1+i$, for some $i$, the reading/writing cost and the distortion are $\frac{N}{\lfloor\frac{N}{2}\rfloor-1+i}$ and $\frac{i}{\lfloor\frac{N}{2}\rfloor-1+i}$, respectively. Since the reading/writing cost monotonically decreases with $i$, and $i$ needs to satisfy $i\leq\frac{\tilde{D}}{1-\tilde{D}}\left(\lfloor\frac{N}{2}\rfloor-1\right)$, the optimum $i^*$ which gives $\ell^*$ is thus $i^*=\frac{\tilde{D}}{1-\tilde{D}}\left(\lfloor\frac{N}{2}\rfloor-1\right)$, which achieves the minimum costs in \eqref{main}. However, in cases where $i^*\notin\mathbb{Z}^+$, we divide all submodels into two sections, assign two separate subpacketizations and apply the scheme on the two sections independently, which achieves the minimum costs in \eqref{main}, after using an optimum ratio for the subsection lengths. To find the optimum ratio, we solve the following optimization problem. Let $\lambda_i$ be the fraction of each submodel with subpacketization $\ell_i=\lfloor\frac{N}{2}\rfloor-1+i$ for some $i=\eta_1,\eta_2\in \mathbb{Z}^+$. Then, based on the average cost and distortion expressions, the minimum reading/writing cost under a given distortion budget is obtained by solving, 
\begin{align}
    \min & \quad \sum_{i=\eta_1,\eta_2}\lambda_i\frac{N}{\lfloor\frac{N}{2}\rfloor-1+i}\nonumber\\
    \text{s.t.} &\quad \sum_{i=\eta_1,\eta_2}\lambda_i\frac{i}{\lfloor\frac{N}{2}\rfloor-1+i}\leq \tilde{D} \nonumber\\
    &\quad \lambda_{\eta_1}+\lambda_{\eta_2}=1\nonumber\\
    &\quad \lambda_{\eta_1},\lambda_{\eta_2} \geq 0.
\end{align}
This problem has multiple solutions that give the same minimum total communication cost. As one of the solutions, consider $\eta_1=0$ and $\eta_2=\eta$, where $\eta=\lceil\frac{\tilde{D}}{1-\tilde{D}}(\lfloor\frac{N}{2}\rfloor-1)\rceil$, 
\begin{align}
    \lambda_0&=1-\frac{\tilde{D}}{\eta}\left(\lfloor\frac{N}{2}\rfloor-1+\eta\right),\label{p0}\\
    \lambda_\eta&=\frac{\tilde{D}}{\eta}\left(\lfloor\frac{N}{2}\rfloor-1+\eta\right)\label{p1}. 
\end{align}
This gives a minimum cost of $C_{\text{min}}=\frac{2}{1-\frac{N}{2}}(1-\tilde{D})$, which matches the terms in \eqref{main}, with $\tilde{D}=\tilde{D}_r$ and $\tilde{D}=\tilde{D}_w$. 

Precisely, for a setting with given $N$, $\tilde{D}_r$ and $\tilde{D}_w$, the reading and writing costs given in \eqref{main} are achievable with corresponding subpacketizations given by,
\begin{align}
    \ell_r^*&\!=\!\!\begin{cases}
    \!\lfloor\frac{N}{2}\rfloor\!-\!1, &\!\!  \text{for $\lambda_0^{[r]}$ of submodel},\\
    \!\lfloor\frac{N}{2}\rfloor\!-\!1\!+\!\lceil\frac{\tilde{D}_r\left(\lfloor\frac{N}{2}\rfloor\!-\!1\right)}{1-\tilde{D}_r}\rceil, &\!\!\! \text{for $1\!\!-\!\!\lambda_0^{[r]}$ of submodel},
    \end{cases}
\end{align}
and
\begin{align}
    \!\ell_w^*&\!=\!\!\begin{cases}
    \!\lfloor\frac{N}{2}\rfloor\!-\!1, &\!\!  \text{for $\lambda_0^{[w]}$ of submodel},\\
    \!\lfloor\frac{N}{2}\rfloor\!-\!1\!\!+\!\!\lceil\frac{\tilde{D}_w\left(\lfloor\frac{N}{2}\rfloor\!-\!1\right)}{1-\tilde{D}_w}\rceil, &\!\!\! \text{for $1\!\!-\!\!\lambda_0^{[w]}$ of submodel},
    \end{cases}
\end{align}
where $\lambda_0^{[r]}$, $\lambda_0^{[w]}$ are $\lambda_0$ in \eqref{p0} with $\tilde{D}$ replaced by $\tilde{D}_r$, $\tilde{D}_w$.

\bibliographystyle{unsrt}
\bibliography{references.asilomar2022}

\end{document}